\documentclass[aip,apl,unsortedaddress,superscriptaddress,reprint]{revtex4-1}
\usepackage{hyperref}
\usepackage{graphicx}
\usepackage{amsmath}
\usepackage{amssymb}
\usepackage{times,mathptmx}
\usepackage{dcolumn}
\usepackage{ifthen}
\usepackage{color}
\usepackage{proof}

\begin{document}
\title[]{Quantifying the spin mixing conductance of EuO/W heterostructures by spin Hall magnetoresistance experiments}

\author{Paul Rosenberger}
\email{p.rosenberger@fz-juelich.de}
\affiliation{Peter Gr\"{u}nberg Institut (PGI-6), Forschungszentrum J\"{u}lich GmbH, D-52428 J\"{u}lich, Germany}
\affiliation{Fakult\"at Physik, Technische Universit\"at Dortmund, 44227 Dortmund, Germany}
\author{Matthias Opel}%
\affiliation{Walther-Mei{\ss}ner-Institut, Bayerische Akademie der Wissenschaften, 85748 Garching, Germany}
\author{Stephan Gepr\"ags}
\affiliation{Walther-Mei{\ss}ner-Institut, Bayerische Akademie der Wissenschaften, 85748 Garching, Germany}
\author{Hans Huebl}
\affiliation{Walther-Mei{\ss}ner-Institut, Bayerische Akademie der Wissenschaften, 85748 Garching, Germany}
\affiliation{Physik-Department, Technische Universit\"at M\"unchen, 85748 Garching, Germany}
\affiliation{Munich Center for Quantum Science and Technology (MCQST), 80799 M\"unchen, Germany}
\author{Rudolf Gross}
\affiliation{Walther-Mei{\ss}ner-Institut, Bayerische Akademie der Wissenschaften, 85748 Garching, Germany}
\affiliation{Physik-Department, Technische Universit\"at M\"unchen, 85748 Garching, Germany}
\affiliation{Munich Center for Quantum Science and Technology (MCQST), 80799 M\"unchen, Germany}
\author{Martina M\"{u}ller}
\affiliation{Department of Physics, University of Konstanz, D-78457 Konstanz, Germany}
\author{Matthias Althammer}
\email{matthias.althammer@wmi.badw.de}
\affiliation{Walther-Mei{\ss}ner-Institut, Bayerische Akademie der Wissenschaften, 85748 Garching, Germany}
\affiliation{Physik-Department, Technische Universit\"at M\"unchen, 85748 Garching, Germany}

\date{\today}

\begin{abstract}
The spin Hall magnetoresistance (SMR) allows to investigate the magnetic textures of magnetically ordered insulators in heterostructures with normal metals by magnetotransport experiments. We here report the observation of the SMR in in-situ prepared ferromagnetic EuO/W thin film bilayers with  magnetically and chemically well-defined interfaces. We characterize the magnetoresistance effects utilizing angle-dependent and field-dependent magnetotransport measurements as a function of temperature. Applying the established SMR model, we derive and quantify the real and imaginary parts of the complex spin mixing interface conductance. We find that the imaginary part is by one order of magnitude larger than the real part. Both decrease with increasing temperature. This reduction is in agreement with thermal fluctuations in the ferromagnet.
\end{abstract}
\maketitle

The spin Hall magnetoresistance opened up the opportunity to study magnetically ordered insulators (MOIs) in bilayer heterostructures with normal metals (NM) via simple magnetotransport experiments~\cite{NakayamaSMR,ChenSMR,Althammer2013,Althammer2018}. Until today, the SMR was successfully employed to probe the magnetic moment configuration in ferrimagnetic, antiferromagnetic, and complex (e.g.~canted or chiral) magnetic phases~\cite{NakayamaSMR,Althammer2013,Aqeel2015,KathrinSMRGIG,Ji2017,Hoogeboom2017,Opel2018,Baldrati2018,Fontcuberta2019,Opel2020,Geprags_2020}. Initially, the role of magnetic sublattices was not considered in the SMR theory framework. However, later studies revealed that the effect is sensitive to the spin texture at the interface, in particular to the arrangement of the  indivdual magnetic sublattices. This enabled studies on spin-canting in ferri- and antiferromagnetic insulators~\cite{KathrinSMRGIG,Ji2017,Hoogeboom2017,Opel2018,Baldrati2018,Opel2020,Geprags_2020}. Interestingly, the SMR in \emph{ferro}magnetic insulators has rarely been investigated, and only recently results on EuS/Pt, EuO/Pt interfaces have been reported~\cite{gomez-perez_strong_2020,mallick_magnetoresistance_2020}. The effect critically depends on the quality  of the MOI/NM interface~\cite{Geprgs2020} and in particular on its transparency for the spin current, described via the complex spin mixing interface conductance $G_{\uparrow\downarrow}=G_\mathrm{r}+i G_\mathrm{i}$. Most importantly, both experiments and theory suggest that in such \emph{ferro}magnetic insulators $G_\mathrm{i}$ exceeds  $G_\mathrm{r}$ as interface compensation effects do not reduce $G_\mathrm{i}$~\cite{gomez-perez_strong_2020,zhang_theory_2019}. This opens up a new regime for SMR experiments as well as for spin-transfer torque effects.

In this letter, we focus on the metastable ferromagnetic semiconductor EuO with a Curie temperature of $69\;\mathrm{K}$. Wefirst present an X-ray photoelectron spectroscopy (XPS) study on both EuO/Pt and EuO/W heterostructures, which reveals very differently defined chemical interfaces. Based on our findings, we focus on the well-defined ferromagnetic EuO/W heterostructure in order to investigate their magnetoresistance (MR) as a function of temperature and applied magnetic field orientation. We identify two contributions to the MR in these samples: (i) SMR originating from the EuO/W interface and (ii) ordinary MR from the metallic W layer. By extracting the SMR parameters, we quantify the real and imaginary parts of the spin mixing interface conductance in the limit of $G_\mathrm{i}>G_\mathrm{r}$. This allows us to investigate the temperature dependence of $G_\mathrm{i}$ and $G_\mathrm{r}$ and find that in agreement with theory it is dominated by spin fluctuations.

Epitaxial and bulk-like EuO thin films with a thickness of about $40\;\mathrm{nm}$ are deposited on single crystalline, (001)-oriented yttria-stabilized zirconia (YSZ) substrates using oxide molecular beam epitaxy~\cite{gerber_thermodynamic_2016, lomker_redox-controlled_2019}.
We applied our preparation process reported in Ref.~\onlinecite{prinz_quantum_2016}, using a deposition temperature of $400\;\mathrm{^\circ C}$.
After confirming the film quality by \textit{in-situ} XPS, RHEED and LEED, we subsequently \textit{in-situ} deposit $4\;\mathrm{nm}$ thin W strips with a width of $250\;\mathrm{\mu m}$ through a shadow mask onto the EuO film via electron beam evaporation. For comparison, we fabricate a reference sample consisting of W-strips with the same thickness deposited directly on a YSZ substrate. To protect the uncovered EuO from oxidation, we deposit a $15\;\mathrm{nm}$ MgO capping layer on all samples via electron beam evaporation. To further confirm the EuO quality, we perform SQUID magnetometry measurments. In order to probe the interface quality of EuO/Pt and EuO/W heterostructures by means of \textit{in-situ} XPS using a Al K$_{\alpha}$ source, we fabricate a second set of samples with plane thin films of $0.3\;\mathrm{nm}$ Pt and $0.4\;\mathrm{nm}$ W, respectively, deposited without shadow mask onto EuO films. For magnetotransport measurements we use a standard 4-point dc measurement technique in a 3D-vector magnet cryostat with a variable temperature insert. For all transport measurements we use a charge current bias of $10\;\mathrm{\mu A}$ and a current reversal method to eliminate any spurious thermal signals in our measurements~\cite{SchreierCSSE,Althammer2018}.


\begin{figure}
\includegraphics[width=\columnwidth]{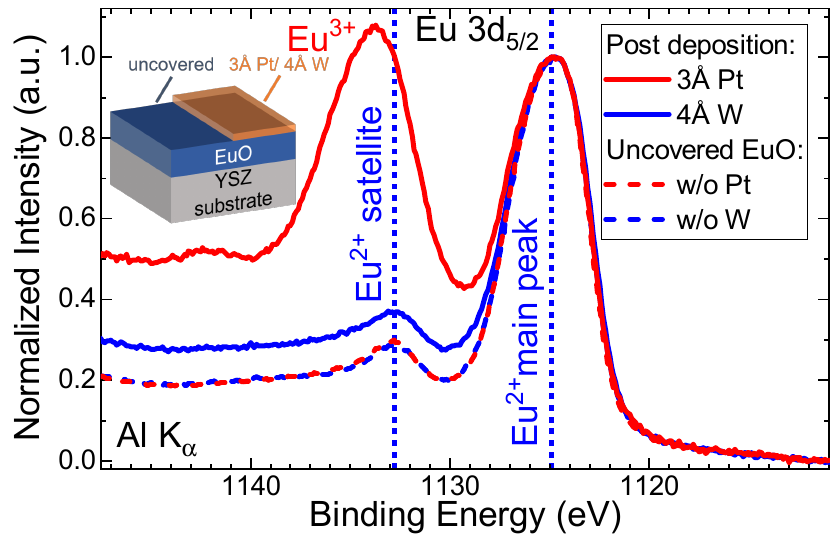}
\caption{Comparison of XPS results for \textit{in-situ} grown EuO/Pt and EuO/W bilayers. For the bare EuO films prior to Pt or W deposition (red and blue dashed lines), the Eu 3d$_{5/2}$ core level exhibits the Eu$^{2+}$ main peak and its satellite at the high binding energy side, indicating ferromagnetic EuO. After deposition of $4\;\mathrm{\AA}$ W the inelastic background is slightly enhanced and only pure Eu$^{2+}$ observed (blue solid line). However, deposition of $3\;\mathrm{\AA}$ Pt results in an intense Eu$^{3+}$ peak, indicating the overoxidation of the EuO/Pt interface towards paramagnetic Eu$_2$O$_3$ or Eu$_3$O$_4$ (red solid line).}
\label{Fig1}
\end{figure}

First, we study the stoichiometry of EuO at the EuO/Pt and EuO/W interfaces by \textit{in-situ} XPS of the Eu 3d$_{5/2}$ core level. To investigate the EuO interface through the metallic Pt and W capping layer, we compare spectra taken before and after metal deposition. In detail, we subtract a constant background and normalize each spectrum to the Eu$^{2+}$ 3d$_{5/2}$ peak height.

The Eu 3d$_{5/2}$ core-level spectra for the two different trilayers are depicted in Fig.~\ref{Fig1}. For the uncovered EuO films (dashed lines), we find only spectral contributions of Eu$^{2+}$, indicating stoichiometric and thus ferromagnetic EuO~\cite{lomker_two-dimensional_2017,lomker_redox-controlled_2019,Mueller2009, caspers_chemical_2011}. After W or Pt deposition (solid lines), the spectra show an enhanced inelastic background at the high binding energy side, which does not hamper the analysis of the EuO composition but simply originates from additional scattering of the photoelectrons while traversing the metal overlayer. For EuO/W, the spectrum in Fig.~\ref{Fig1} also exhibits a pronounced Eu$^{2+}$ main peak and a satellite at its higher binding energy side, i.e.~the signature of pure Eu$^{2+}$, indicating ferromagnetic EuO at the well-defined EuO/W interface. 

For EuO/Pt, in contrast, we find an intense Eu$^{3+}$ peak at an even higher binding energy compared to the Eu$^{2+}$ satellite \cite{caspers_chemical_2011}. It is thus evident that upon deposition of Pt onto a stoichiometric EuO film an interfacial oxidation process of the initially ferromagnetic EuO takes place, causing the formation of paramagnetic Eu$_2$O$_3$ or Eu$_3$O$_4$~\cite{gerber_thermodynamic_2016}. Thus, no sharp interface between a magnetically ordered insulator and a normal metal persists for this heterostructure. Hence, we focus on EuO/W heterostructures in the following.


\begin{figure}
\includegraphics[width=85mm]{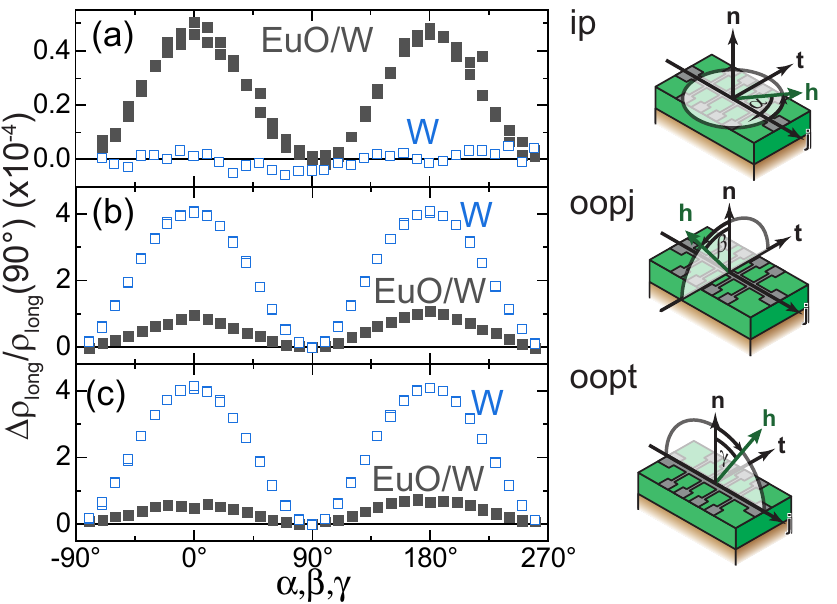}
\caption{Angle-dependent magnetotransport at $T=15\;\mathrm{K}$ and $\mu_0H=2\;\mathrm{T}$ for a EuO/W heterostructure (black squares) and a bare W film (open blue squares) utilizing (a) in-plane (ip) rotation plane, (b) out-of-plane perpendicular to $\mathbf{j}$ (oopj) rotation plane, and (c) out-of-plane perpendicular to $\mathbf{t}$ (oopt) rotation plane. The illustrations on the right show rotation plane of the external magnetic field with respect to the sample and current direction. For EuO/W, we observe in all rotation planes an angle-dependence of $\rho_\mathrm{long}$, consistent with the combined action of SMR and ordinary MR. For the bare W layer, there is no angle-dependence visible in the ip rotation confirming that the EuO/W interface is necessary to observe the SMR signature.}
\label{Fig2}
\end{figure}

For the investigation of the transport properties, we first perform angle-dependent magnetoresistance (ADMR) measurements. The orientation of an external magnetic field $\mathbf{h}$ with constant magnitude $\mu_0 H$ is varied in three distinct rotation planes while the longitudinal resistivity $\rho_\mathrm{long}$ of the sample is determined as a function of the rotation angle. The current direction $\mathbf{j}$ and the surface normal $\mathbf{n}$ define our orthogonal coordinate system via the transverse direction $\mathbf{t}=\mathbf{n}\times\mathbf{j}$ (see illustration in Fig.~\ref{Fig2}). Utilizing these three unit vectors, we define three orthogonal rotation planes for the magnetic field (referred to as ip, oopj, and oopt, see Fig.~\ref{Fig2}). Using all three rotation planes we analyze the symmetry of the observed magnetoresistance to determine its origin. We calculate the longitudinal magnetoresistance $MR(\phi)= [\rho_\mathrm{long}(\phi)/\rho_\mathrm{long}(\phi=90^\circ)]-1$ for $\phi=\alpha, \beta, \gamma$. In Fig.~\ref{Fig2}, we show the obtained results for the bare W reference (blue open squares) and the EuO/W sample (black filled squares) at $T=15\;\mathrm{K}$ and $\mu_0H=2\;\mathrm{T}$.

\begin{figure}
\includegraphics[width=85mm]{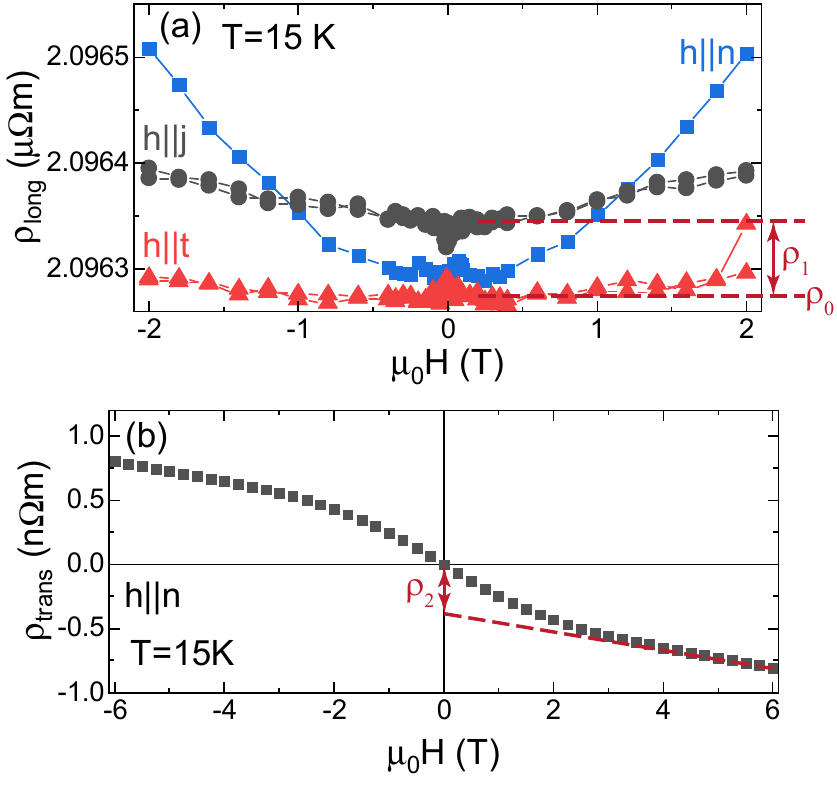}
\caption{Field-dependent magnetotransport for the EuO/W sample at $T=15\;\mathrm{K}$: (a) $\rho_\mathrm{long}$ as a function of the applied magnetic field for $\mathbf{h}$ oriented along $\mathbf{j}$ (black dots), $\mathbf{t}$ (red triangles), and $\mathbf{n}$ (blue squares). For $\mathbf{h}\parallel\mathbf{n}$, we observe a quadratic field dependence, typical for the OMR in W. (b) $\rho_\mathrm{trans}$ (black symbols) as a function of the applied magnetic field for $\mathbf{h}\parallel\mathbf{n}$. At large magnetic fields a linear field dependence (dashed red line) originating from the ordinary Hall effect is observed. The red dashed lines and arrows in both panels illustrate the extraction of $\rho_0$, $\rho_1$ and $\rho_2$ from these measurements.}
\label{Fig3}
\end{figure}

For the bare W reference sample, we only observe an angle-dependence for the oopj and oopt rotations with a $\cos^2$-dependence and a maximum $MR=4\times10^{-4}$  for $\mathbf{h}\parallel\mathbf{n}$. Such a behaviour is typical for an ordinary magnetoresistance (OMR) where the resistance of the sample increases if the magnetic field is applied perpendicular to the sample plane. In our samples, this OMR exhibits a quadratic magnetic field dependence (see Fig.~\ref{Fig3}(a)). For the ip rotation plane, we do not find any systematic angle-dependence within the resolution ($MR<5\times10^{-6}$) of our setup (Fig.~\ref{Fig2}a). 

For our EuO/W sample, in contrast, we observe an angle-dependence for the ip rotation, following a $\cos^2$ behavior and showing a maximum of $MR=5\times10^{-5}$ at $\mathbf{h}\parallel\mathbf{j}$. This angle-dependence and position of the maximum (i.e.~its phase) is consistent with the SMR for ip rotations in ferrimagnetic insulators where the magnetization follows the external magnetic field~\cite{NakayamaSMR,Althammer2013,ChenSMR,Geprags_2020}. In case of the oopj rotation, the MR exhibits a $\cos^2$ dependence and maximum of $MR=11\times10^{-5}$ for $\mathbf{h}\parallel\mathbf{n}$. For this rotation plane, both SMR and OMR of the W layer contribute to the MR. In the oopt rotation plane, where only the OMR contributes, we observe a $\cos^2$ angle-dependence and a maximum of $MR=7\times10^{-5}$. The fact that the sum of the maximum MR for ip and oopt rotation planes nicely matches the maximum MR in the oopj rotation, confirms our assumption that two effects contribute to the MR in the EuO/W sample: the SMR of the EuO/W interface and the OMR of the W layer. Moreover, since we do not observe any significant angle-dependence for the ip rotation in the W reference sample, we conclude that the EuO/W interface is crucial to observe the SMR effect. The SMR symmetry is attributed to spin current transfer across the EuO/W interface governed by the spin mixing conductance. In addition, we can rule out the Hanle MR~\cite{Velez_2016, Dyakonov_2007} as a dominant contribution in our samples. 

To complement the picture of the angle-dependent MR in our EuO/W sample, we also investigate the evolution of $\rho_\mathrm{long}$ as a function of the magnitude of the applied magnetic field $\mu_0 H$ for three fixed field orientations: $\mathbf{h} \parallel \mathbf{j}$, $\mathbf{h} \parallel \mathbf{t}$, and $\mathbf{h} \parallel \mathbf{n}$ (see illustration in Fig.~\ref{Fig2} for a definition of these orientations). The results are plotted in Fig.~\ref{Fig3}(a) for $T=15\;\mathrm{K}$.
Obviously, $\rho_\mathrm{long}$ is larger for $\mathbf{h} \parallel \mathbf{j}$ than for $\mathbf{h} \parallel \mathbf{t}$ which is fully consistent with the SMR nature of the in-plane MR~\cite{NakayamaSMR,Althammer2013,ChenSMR}. For these two field orientations, moreover, $\rho_\mathrm{long}$ increases gradually with increasing magnetic field magnitude. This weak magnetic field-dependence may either originate from a small Hanle MR~\cite{Velez_2016,Dyakonov_2007} only present in our EuO/W bilayer and not in the W reference sample, or a slight misalignment of the magnetic field with respect to the sample plane, such that the OMR sensitive to the out-of-plane magnetic field component also contributes in our measurements. For $\mathbf{h} \parallel \mathbf{n}$, $\rho_\mathrm{long}$ exhibits a pronounced, parabola-shaped magnetic field-dependence, as expected for an OMR in the W layer.

To complete the picture, we plot the magnetic field-dependence of the transverse resistivity $\rho_\mathrm{trans}$ for $\mathbf{h} \parallel \mathbf{n}$ in Fig.~\ref{Fig3}(b) at $15\;\mathrm{K}$ for the EuO/W sample. For large external magnetic fields, we observe a linear increase with negative slope in $\rho_\mathrm{trans}$ originating from the ordinary Hall effect in the W layer (dashed red line in Fig.~\ref{Fig3}(b)). For small magnetic fields $\mu_0 H \leq 3 \;\mathrm{T}$, $\rho_\mathrm{trans}$ deviates from this linear dependence. We attribute this deviation to the SMR, which causes an anomalous Hall-like contribution to $\rho_\mathrm{trans}$~\cite{ChenSMR,Meyer_SHAHE_2014}. At these small fields, the magnetization of the EuO layer is no longer saturated along $\mathbf{n}$, but starts to align in the sample plane due to the increasing influence of shape anisotropy.

We carried out field-dependent MR experiments as a function of temperature (not shown here) for all three field orientations to further investigate the temperature-dependence of the SMR in our EuO/W sample. To quantify the SMR from these measurements, we utilize the SMR expression for $\rho_\mathrm{long}$ from Ref.~\onlinecite{ChenSMR}
\begin{equation}
\rho_\mathrm{long}=\rho_0+\rho_1\,(1-m_\mathrm{t}^2)\;,
\label{eq:SMR:long}
\end{equation}
with $m_\mathrm{t}$ the projection of the magnetization orientation $\mathbf{m}$ in EuO onto the $\mathbf{t}$-direction. From the field dependent MR measurements for $\mathbf{h} \parallel \mathbf{j}$ ($m_\mathrm{t}=0$) and $\mathbf{h} \parallel \mathbf{t}$ ($m_\mathrm{t}=1$), we extract $\rho_0$ and $\rho_1$ as illustrated in Fig.~\ref{Fig3}(a) for $\mu_0 H=250\;\mathrm{mT}$, where the magnetization in the EuO is already saturated within the sample plane and the influence of the magnetic field-dependence of $\rho_\mathrm{long}$ is negligible. For $\rho_\mathrm{trans}$, we write
\begin{equation}
\rho_\mathrm{trans}=\rho_2 m_\mathrm{n}+\rho_3\,m_\mathrm{j}m_\mathrm{t}\;,
\label{eq:SMR:trans}
\end{equation}
in the framework of the SMR theory~\cite{ChenSMR} with $m_\mathrm{n}$ and $m_\mathrm{t}$ the projections of $\mathbf{m}$ onto $\mathbf{n}$ and $\mathbf{j}$, respectively. For $\mathbf{h} \parallel \mathbf{n}$ ($m_\mathrm{j},m_\mathrm{t}=0$), we extract $\rho_2$ by linear fitting the $\rho_\mathrm{trans}$ data for $|\mu_0 H|>4\;\mathrm{T}$ and extrapolating to zero magnetic field as indicated in Fig.~\ref{Fig3}(b). From the linear fit we also extract the ordinary Hall constant for our EuO/W bilayer and find $-8\times10^{-11}\;\mathrm{\Omega m T^{-1}}$ in agreement with Hall measurements on W in previous studies~\cite{hao_beta_2015}. 

\begin{figure}
\includegraphics[width=85mm]{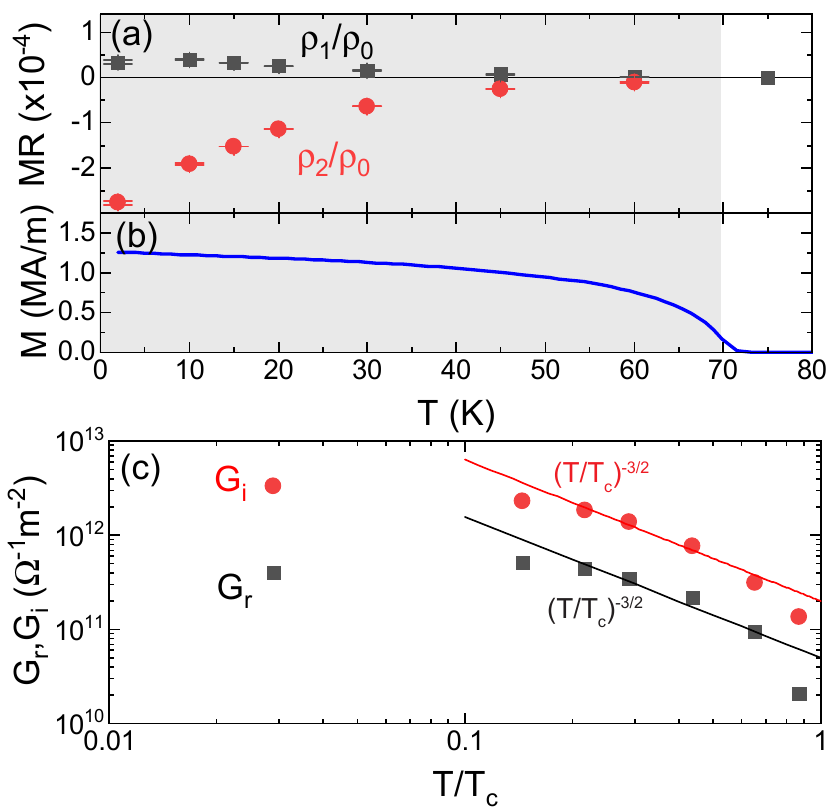}
\caption{(a) Temperature dependence of $\rho_1/\rho_0$ and $\rho_2/\rho_0$ extracted from the field-dependent magnetoresistance measurements on the EuO/W sample. (b) Magnetization versus temperature determined from SQUID magnetometry after field cooling the sample and measuring at zero magnetic field. (c) Extracted values for $G_\mathrm{r}$ (black squares) and $G_\mathrm{i}$ (red circles) as a function of the reduced temperature $T/T_\mathrm{c}$. The black and red line indicate a $(T/T_\mathrm{c})^{-3/2}$-dependence, which agrees well with the extracted temperature dependence for $G_\mathrm{r}$ and $G_\mathrm{i}$ in the intermediate $T/T_\mathrm{c}$-range.}
\label{Fig4}
\end{figure}

The results for $\rho_1/\rho_0$ and $\rho_2/\rho_0$ are plotted as a function of temperature in Fig.~\ref{Fig4}(a). For both SMR contributions, we observe an increase with decreasing $T$. The values for $\rho_1/\rho_0$ seem to saturate at low temperatures. Interestingly, both contributions vanish at $75\;\mathrm{K}$, just above the Curie temperature $T_c\approx69\;\mathrm{K}$ determined from SQUID magnetometry measurements (see Fig.~\ref{Fig4}(b)), indicated by the grey shaded area in Figs.~\ref{Fig4}(a),(b). Within our experimental resolution, we do not find any evidence that the SMR persists above $T_\mathrm{c}$, as for example observed in Cr$_2$O$_3$/Pt bilayers~\cite{Schlitz2018}. As discussed further below this may originate from the significant reduction of $G_\mathrm{r}$ and $G_\mathrm{i}$ due to the localized $4f$-electrons contributing to the SMR for Eu$^{2+}$ as opposed to the $3d$-electrons in Cr$^{3+}$. For all investigated temperatures, the absolute value of $\rho_2/\rho_0$ is larger than $\rho_1/\rho_0$, which we attribute to the facts that $G_\mathrm{i}>G_\mathrm{r}$ for all temperatures (as quantified below) and $\rho_2/\rho_0$ is dominated by $G_\mathrm{i}$ whereas $\rho_1/\rho_0$ is dominated by $G_\mathrm{r}$. This finding is exactly opposite to the results obtained in ferrimagnetic YIG/Pt heterostructures with in-situ interfaces, where $\rho_1/\rho_0$ is always much larger than $\rho_2/\rho_0$  and compensation effects significantly reduce $G_\mathrm{i}$.~\cite{Althammer2013,SibylleSMR,zhang_theory_2019}


We now use the extracted temperature-dependent $\rho_1/\rho_0$ and $\rho_2/\rho$ to extract values for the complex spin mixing conductance $G_{\uparrow\downarrow}=G_\mathrm{r}+iG_\mathrm{i}$ of the EuO/W interface.
From SMR theory~\cite{ChenSMR}, we get the following expression for $\rho_1/\rho_0$
\begin{equation}
    \frac{\rho_1}{\rho_0}=\frac{\theta_\mathrm{SH}^2}{t_\mathrm{NM}}\lambda_\mathrm{sf}\mathrm{Re}\left[\frac{2\rho_\mathrm{NM} \lambda_\mathrm{sf} G_{\uparrow\downarrow} \tanh^2(\frac{t_\mathrm{NM}}{2\lambda_\mathrm{sf}})} {1+2\rho_\mathrm{NM} \lambda_\mathrm{sf} G_{\uparrow\downarrow} \coth(\frac{t_\mathrm{NM}}{\lambda_\mathrm{sf}})} \right]\,.
    \label{eq:SMR:rho1}
\end{equation}
with $\theta_\mathrm{SH}$ the spin Hall angle, $\lambda_\mathrm{sf}$ the spin diffusion length, $\rho_\mathrm{NM}$ the resistivity and $t_\mathrm{NM}$ the thickness of the NM. For $\rho_2/\rho_0$, we write down the following expression from SMR theory.
\begin{equation}
    \frac{\rho_2}{\rho_0}=-\frac{\theta_\mathrm{SH}^2}{t_\mathrm{NM}}\lambda_\mathrm{sf}\mathrm{Im}\left[\frac{2\rho_\mathrm{NM} \lambda_\mathrm{sf} G_{\uparrow\downarrow} \tanh^2(\frac{t_\mathrm{NM}}{2\lambda_\mathrm{sf}})} {1+2\rho_\mathrm{NM} \lambda_\mathrm{sf} G_{\uparrow\downarrow} \coth(\frac{t_\mathrm{NM}}{\lambda_\mathrm{sf}})} \right]\,.
    \label{eq:SMR:rho2}
\end{equation}
It is important to note that in most previous reports~\cite{ChenSMR} the above expressions were approximated for the case of $G_\mathrm{r} \gg G_\mathrm{i}$, a valid assumption for ferrimagnetic and antiferromagnetic insulators, were compensation effects~\cite{zhang_theory_2019} can drastically reduce $G_\mathrm{i}$. For ferromagnetic insulators we need to account for $G_\mathrm{r}$ and $G_\mathrm{i}$, which especially leads to contributions of $G_\mathrm{i}$ to $\rho_1/\rho_0$, mostly neglected for ferri- and antiferromagnetic insulators. Assuming $\lambda_\mathrm{sf} \rho_\mathrm{NM} G_\mathrm{r} \ll 1$ (spin transparency of the interface far away from the ideal limit), we obtain the following approximation for $G_\mathrm{i}$ by combining Eqs.(\ref{eq:SMR:rho1}) and (\ref{eq:SMR:rho2})
\begin{equation}
G_\mathrm{i}\approx \frac{\rho_2\rho_0^{-1}-\left[\rho_1\rho_0^{-1}\rho_2\rho_0^{-1} \left[ \theta_\mathrm{SH}^2 t_\mathrm{NM}^{-1} \lambda_\mathrm{sf}^2 \rho_0 \tanh\left(\frac{t_\mathrm{NM}}{2\lambda_\mathrm{sf}}\right) \right]^{-1}\right]}{-\theta_\mathrm{SH}^2 t_\mathrm{NM}^{-1} \lambda_\mathrm{sf}^2 \rho_0 \tanh\left(\frac{t_\mathrm{NM}}{2\lambda_\mathrm{sf}}\right)}\,.
\label{eq:Gi}
\end{equation}
Within this approximation, we have set $\rho_\mathrm{NM}=\rho_0$, as $\rho_0$ is determined from the field-dependent MR measurements. $G_\mathrm{r}$ can then be determined from Eq.(\ref{eq:SMR:rho1}) and inserting the approximated $G_\mathrm{i}$. We obtain a quadratic equation for $G_\mathrm{r}$, such that two solutions for $G_\mathrm{r}$ exist. We select the solution with the correct temperature dependence of the absolute value of $G_\mathrm{r}$ (decreasing with increasing temperature). To quantify $G_\mathrm{r}$ and $G_\mathrm{i}$, we assume temperature-independent $\lambda_\mathrm{sf}=1.5\;\mathrm{nm}$ and $\theta_\mathrm{SH}=0.2$ for W taking into account the mixed phase of $\alpha$- and $\beta$-W in our sample~\cite{Pai2012}. Not accounting for a temperature dependence of these two parameters is justified by the weak temperature dependence of $\rho$ of the W layer, as $\theta_\mathrm{SH}$ and $\lambda_\mathrm{sf}$ depend on the NM resistivity~\cite{Sagasta2016,Sagasta2018}. For our EuO/W sample, $\rho$ only changes by $1.4\times10^{-2}$ in the investigated temperature range. We want to emphasize that a change of the assumed values for $\lambda_\mathrm{sf}$ and $\theta_\mathrm{SH}$ will only influence the absolute numbers of $G_\mathrm{r}$ and $G_\mathrm{i}$, but does not affect the general temperature dependence for the two parameters. The results of this procedure are shown in Fig.~\ref{Fig4}(c) as a function of the reduced temperature $T/T_\mathrm{c}$.

For low temperatures, we find $G_\mathrm{r}=4\times10^{11}\;\mathrm{\Omega^{-1}m^{-2}}$ and  $G_\mathrm{i}=3\times10^{12}\;\mathrm{\Omega^{-1}m^{-2}}$. The obtained values agree well with results on EuS/Pt~\cite{gomez-perez_strong_2020}. Compared to YIG/Pt interfaces~\cite{Althammer2013,Weiler2013,SibylleSMR}, however, the extracted $G_\mathrm{r}$ is significantly reduced for the EuO/W interface indicating a lower coupling strength between the mobile electrons in the NM and the localized electrons of the magnetic moments in EuO. We attribute this observation to the different orbital nature of the localized electrons. For Eu$^{2+}$ the localized electrons are in 4f-orbitals, while for Fe$^{3+}$ they are located in 3d-orbitals. Thus, we assume that coupling effects via orbital overlap are reduced for EuO as compared to YIG, which reasonably explains the differences between YIG/NM and EuO/NM interfaces.

For $G_\mathrm{r}$ and $G_\mathrm{i}$ we observe a monotonous decrease with increasing temperature. The temperature-dependence of $G_\mathrm{r}$ and $G_\mathrm{i}$ is explained by the influence of thermal fluctuations in the magnetic lattice of the MOI~\cite{Bender2015,Wang2018,zhang_theory_2019}. In this way, we can describe the temperature dependence as $G_\mathrm{r}\approx(1-2n/s)G_\mathrm{r}(T=0)$ and $G_\mathrm{i}\approx(1-n/s) G_\mathrm{i}(T=0)$, where $n$ is the density of magnons per unit cell and $s$ is the saturation magnetization per unit cell in units of $\hbar$. 
In the limit of a three dimensional ferromagnet with parabolic magnon band dispersion and an intermediate $T/T_\mathrm{c}$-range, both $G_\mathrm{r}$ and $G_\mathrm{i}$ both should follow a $(T/T_\mathrm{c})^{-3/2}$ law, i.e.~they scale inversely as the number of magnons. Indeed for $0.1<T/T_\mathrm{c}<0.7$, the extracted $G_\mathrm{r}$ and $G_\mathrm{i}$ agree well with a $(T/T_\mathrm{c})^{-3/2}$ dependence (black and red line in Fig.~\ref{Fig4}(c)), further confirming the theoretical model.

In summary, we present a detailed study of the MR in EuO/W bilayers. Initial XPS studies on both EuO/Pt and EuO/W heterostructures revealed a well-defined interface only for the EuO/W  system. In this sample, we find two contributions to the observed MR: SMR and ordinary MR. We utilize the SMR contribution and its temperature dependence to quantify $G_\mathrm{r}$ and $G_\mathrm{i}$. From this analysis we find $G_\mathrm{i}>G_\mathrm{r}$ for our EuO/W sample consistent with recent experiments on EuS/Pt interfaces. Our results confirm that in ferromagnetic insulators a new regime for spin-transfer torque experiments can be established, where the field-like symmetry is the dominant contribution. Moreover, our approach to quantify $G_\mathrm{r}$ and $G_\mathrm{i}$ provides an alternative approach to investigate the spin mixing conductance, if the spin transport parameters $\theta_\mathrm{SH}$ and $\lambda_\mathrm{sf}$ and their temperature dependence of the NM are well known.


\begin{acknowledgements}
We acknowledge financial support by the German Research Foundation (Deutsche Forschungsgemeinschaft, DFG) via Germany's Excellence
Strategy -- EXC-2111 -- 390814868 and project AL 2110/2-1. P.R. and M.M. acknowledged financial support by the International Collaborative Research Center TRR160 (Project C9). 
\end{acknowledgements}

\section*{AIP Publishing Data Sharing Policy}
The data that support the findings of this study are available from the corresponding author upon reasonable request.

\bibliography{Bibliography}
\end{document}